\documentstyle[12pt]{article}
 \newcommand{\be}{\begin{equation}}
\newcommand{\ee}{\end{equation}}
\title{HOT NONPERTURBATIVE  QCD\thanks{Lecture at the International
School of Physics "Enrico Fermi", Varenna, 27 June -- 7 July 1995} }
\author{ Yu.A.Simonov\\ Institute of Theoretical and Experimental
Physics\\ Moscow, 117259} \date{}
\begin{document}
 \maketitle

\begin{abstract}

Starting from the new background field formalism for $T>0$, with
nonperturbative (NP) background given as gauge invariant fiel
correlators, new perturbation theory and diagram technic is
introduced.

 Confined and deconfined phases are explicitly
 described and critical temperature $T_c$ is
 expressed in terms of the scale anomaly term.
 Resulting numerical estimates of $T_c$ agree
 well with lattice data. Spacial area law is
 shown to follow naturally for $T>T_c$, and a
 set of Hamiltonians is obtained for screening
 masses of mesons and glueballa depending on the
 Matsubara frequences.

 Thelowest screening masses and wave functions
 of mesons obtained agree with lattice data and
 earlier calculations. Screening glueball masses
 and wave functions are also  computed from the
 Hamiltonian and two different regimes are
 observed in high and low $T$  regions.

 The infrared catastrophe of hot QCD is shown to
 be cured by the NP background.
 \end{abstract}

 \section{Introduction}

The finite temperature QCD provides
a unique insight into the structure
	of the QCD vacuum, in particular it
gives an important information about
the mechanism of
confinement. It is also of more practical
interest, since the nature of the  deconfinement transition has
its bearing on the cosmology  and the hot
QCD plasma can possibly  be tested in heavy ion collisions.

{}From  theoretical point of view the  hot
QCD is a unique theoretical laboratory where both perturbative (P)
 and  nonperturbative (NP) methods can be
applied in different temperature regimes.

It is believed that the perturbative QCD is applicable in the deconfined
phase at large enough temperatures $T$, where the effective coupling
 constant $g(T)$ is small [1], while at small $T$ (in the confined phase)
 the NP effects instead are most important. However
 even at large T  the physics is not that simple: some effects, like
 screening (electric gluon mass), need a resummation of the
 perturbative series [2], while the effects connected with the
 magnetic gluon mass demonstrate the infrared divergence of the
 series [3].

During last years there appeared a lot of lattice data which point
at the NP character of dynamics above $T_c$. Here
belong i) area law of spacial Wilson loops [4] ii) screening masses
of mesons and baryons [5] and glueballs [6] iii) temperature
dependence of Polyakov--line correlators [7].

In addition, behaviour of $\varepsilon-3p$ above $T_c$ has a bump
incompatible with the simple quark--gluon  gas picture [8].

Thus the inclusion of NP configurations into
QCD at $T>0$ and also at $T>T_c$ is necessary.

Recently a systematic method for QCD was developed, treating $NP$
fields as a background and doing perturbative expansion around that
both for $T=0$ [9] and $T>0$ [10].

To describe the phase transition a simple choice of deconfined phase
was suggested where all  $NP$ color magnetic configurations are
kept intact as in the confined phase, whereas colorelectric
correlators responsible for confinement, vanish.

This picture  together with  the background perturbation theory
 forms a basis of quantitative calculations, where field
correlators (condensates) are used as the $NP$ input.

The plan  of
the lecture is as follows. In the second chapter the new
background field formalism is presented, based on the familiar background
field method [11] modified for $T>0$ and incorporating
the 'tHooft's identify for
integration  over quantal and background fields.

The temperature  phase transition is discussed in chapter 3 and resulting
predictions for $T_c$  are compared with lattice data.

  In chapter 4 the spacial Wilson loops are computed both below and above
$T_c$;
 it is shown  how at large $T$ a new regime -- that of dimensional reduction --
 appears and the spacial string tension is discussed.

Chapter 5 is devoted to screening masses of mesons, and glueballs,
which were  considered recently in [5,6].

 In chapter 6 the analysis is given of the
 infrared catastrophe of the hot QCD;
it is shown,   that  it is cured
naturally in the present formalism by nonperturbative
 contributions.

Some prospectives of the method and
discussion of results are given in Conclusions.

\section{New background field formalism}

 We derive here basic formulas for the partition function, free energy and
Green's  function in the NP background formalism at
$T>0$ [10]. The total gluonic field $A_{\mu}$ is split into a
perturbative part $a_{\mu}$ and NP background $B_{\mu}$
\be
A_{\mu}=B_{\mu}+a_{\mu}
\ee
where both $B_{\mu}$ and  $a_{\mu}$ are subject to periodic boundary
conditions. The principle of this separation is immaterial for our purposes
here, and one can average over fields $B_{\mu}$ and $a_{\mu}$ independently
using the 'tHooft's identity\footnote{private communication to one of the
authors (Yu.S.), December 1993.}
\be
 Z=\int
DA_{\mu} exp (-S(A)) = \frac{\int DB_{\mu}\eta(B)\int Da_{\mu} exp
(-S(B+a))}{\int DB_{\mu}\eta(B)}
  \ee
  $$ \equiv<<exp(-S(B+a)>_a>_B $$
   with
arbitrary weight $\eta(B)$. In our case we choose $\eta(B)$ to fix field
correlators and string tension at their observed values.

The partition function can be written as $$ Z(V,T,\mu=0) =<Z(B)>_B\;,$$
\begin{equation}
Z(B)=N\int D\phi exp (-\int^{\beta}_0 d\tau \int d^3x L(x,\tau))
\end{equation}
 where $\phi$ denotes all set of fields $a_{\mu}, \Psi, \Psi^+,N$ is a
normalization  constant, and the sign $<>_B$  means some averaging over
(nonperturbative) background fields $B_{\mu}$,
 as in (2)
. Furthermore, we have
$$L(x,\tau)=\sum^{8}_{i=1} L_i,$$
where
\begin{eqnarray}
\nonumber
L_1=\frac{1}{4} (F^a_{\mu\nu}(B))^2;  L_2=\frac{1}{2} a_{\mu}^a
W_{\mu\nu}^{ab} a_{\nu}^b,
\\
L_3=\bar{\Theta}^a (D^2(B))_{ab}\Theta^b; L_4=-ig\bar{\Theta}^a (D_{\mu},
a_{\mu})_{ab}\Theta^b
\\
\nonumber
L_5=\frac{1}{2}\alpha (D_{\mu}(B)a_{\mu})^2;  L_6=L_{int} (a^3,a^4)
\\
\nonumber
L_7=- a_{\nu} D_{\mu}(B) F_{\mu\nu}(B); L_8=\Psi^+(m+\hat{D}(B+a))\Psi
\end{eqnarray}

Here $\bar{\Theta},\Theta$
are ghost fields, $\alpha$- gauge--fixing constant, $L_6$
contains 3--gluon-- and 4--gluon vertices, and we keep the most general
background
field $B_{\mu}$, not satisfying classical equations, hence the
 appearance of $L_7$.

 The inverse gluon propagator in the background gauge is
\begin{equation}
W^{ab}_{\mu\nu} =- D^2(B)_{ab} \cdot \delta_{\mu\nu} - 2 g F^c_{\mu\nu}(B)
f^{acb}
\end{equation}
where
\begin{equation}
(D_{\lambda})_{ca} = \partial_{\lambda} \delta_{ca} -
ig T^b_{ca} B^b_{\lambda}
\equiv
\partial_{\lambda} \delta_{ca} - g f_{bca} B^b_{\lambda}
\end{equation}

We consider first the case of pure  gluodinamics, $L_8\equiv 0$.

Integration over ghost and gluon degrees of freedom in (3) yields
\begin{eqnarray}
\nonumber
Z(B) =N'(det W(B))^{-1/2}_{reg} [det (-D_{\mu}(B)
D_{\mu}(B+a))]_{a=\frac{\delta}{\delta J}} \times
\\
\times \{ 1+ \sum^{\infty}_{l=1} \frac{S_{int}^l}{l!} (a= \frac{\delta}{\delta
J}) \}
exp (-\frac{1}{2} J W^{-1}J)_{J_{\mu}= \;\;\;\;\;D_{\mu}(B)F_{\mu\nu}(B)}
\end{eqnarray}

One can consider  strong background fields, so that $gB_{\mu}$ is large (as
compared to $\Lambda^2_{QCD}$), while $\alpha_s=\frac{g^2}{4\pi}$
in that strong background is small at all distances [9].

In this case Eq. (7) is a perturbative sum in powers of $g^n$,
arising from expansion in $(ga_{\mu})^n$.

In what follows we shall discuss the Feynman graphs for the free energy
$F(T),$
connected to $Z(B)$ via
\begin{eqnarray}
F(T) = -T ln <Z(B)>_B
\end{eqnarray}

As will be seen, the lowest  order graphs already  contain  a nontrivial
dynamical mechanism for  the deconfinement transition, and those will be
considered in the next section.

To the lowest order in $ga_{\mu}$ the partition function and free
energy are $$ Z_0=<exp(-F_0(B)/T)>_B, $$ \be F_0(B)/T=\frac{1}{2} ln
det W-lndet (-D^2(B))= \ee $$
=Sp\int^{\infty}_0\zeta(t)\frac{dt}{t}(-\frac{1}{2}e^{-tW}+e^{tD^2(B)})
$$
where $\hat{W}=-D^2(B)-2g\hat{F}$ and $D^2(B)$ is the inverse gluon and
ghost propagator  respectively, $\zeta(t)$ is a regularizing factor
[10].

The ghost propagator can be written as [10],[12]
\be
(-D^2)^{-1}_{xy}=<x|\int^{\infty}_0 dt e^{tD^2(B)}|y>=
\int^{\infty}_0dt(Dz)^w_{xy}e^{-K}\Phi(x,y)
\ee
where standard notations [12] are used
$$K=\frac{1}{4}\int^s_0d\lambda\dot{z}^2_{\mu}~, ~~\Phi(x,y)=P exp
ig\int^x_yB_{\mu}dz_{\mu}$$
and a winding path integral is introduced [10]
\be
(Dz)^w_{xy}=\lim_{N\to
\infty}\prod^{N}_{m=1}\frac{d^4\zeta(m)}{(4\pi\varepsilon)^2}
\sum^{\infty}_{n=0,\pm1,..}
\int\frac{d^4p}{(2\pi)^4}e^{ip(\sum\zeta(m)-(x-y)-n\beta\delta_{\mu 4})}
\ee
with $\beta=1/T$. For the gluon propagator an analogous expression holds
true, expect that in (4) one should insert gluon spin factor $P_F exp
2g\hat{F}$ inside $\Phi(x,y)$. For a quark propagator the sum over windings
in (5) acquires the factor $(-1)^n$ and quark spin factor is $exp
g\sigma_{\mu\nu}F_{\mu\nu}$ [10].

\section{The temperature phase transition in QCD}

We are now in position to make expansion of $Z$ and $F$ in powers of
$ga_{\mu}$ (i.e. perturbative expansion in $\alpha_s$), and the
leading--nonperturbative term $Z_0, F_0$ -- can be represented as a sum of
contributions with different $N_c$ behaviour of which we systematically will
keep the leading terms $0(N_c^2),0(N_c)$ and $0(N_c^0)$.

To describe the temperature phase transition one should specify
phases and compute free energy. For the confining phase to lowest
order in $\alpha_s$ free energy is given by Eq.(3) plus contribution
of energy density $\varepsilon $ at zero temperature
\be
F(1)=\varepsilon V_3-\frac{\pi^2}{30}V_3T^4-T\sum_s\frac{V_3(2m_s
T)^{3/2}}{8\pi^{3/2}}e^{-m_{s/T}}+0(1/N_c)
\ee
where $\varepsilon$ is defined by scale anomaly [13]
\be
\varepsilon \simeq
-\frac{11}{3}N_c\frac{\alpha_s}{32\pi}<(F^a_{\mu\nu}(B))^2>
\ee
and the next terms in (12) correspond to the contribution of mesons (we
keep only pion gas) and glueballs. Note that $\varepsilon=0(N^2_c)$
while two other terms in (12) are $0(N^0_c)$.

For the second phase (to be the high temperature phase) we make an
assumption that there all color magnetic field correlators are the same as
in the first phase, while all color electric fields vanish. Since at
$T=0$ color--magnetic correlators (CMC) and color--electric
correlators (CEC) are equal due to the Euclidean $0(4)$ invariance,
one has
\be
<(F^a_{\mu\nu}(B))^2>=<(F^a_{\mu\nu})^2>_{el}+<(F^a_{\mu\nu})^2>_{magn};
<F^2>_{magn}=<F^2>_{el}
\ee

The string tension $\sigma$ which characterizes confinement is due to the
electric fields [14], e.g. in the plane (i4)
\be
\sigma=\sigma_E=\frac{g^2}{2}\int\int
d^2x<trE_i(x)\Phi(x,0)E_i(0)\phi(0,x)>+...
\ee
where dots imply higher order terms in $E_i$.

Vanishing of $\sigma_E$ liberates gluons and quarks, which will contribute
to the free energy in the deconfined phase their closed loop terms
(10) with all possible windings. The CMC enter via perimeter
contribution\\ $<\Phi(x,x)>\equiv \Omega$ (see (9,10)).  As a result one
has for the high-temperature phase (phase 2) (cf.[10]).
 \be
F(2)=\frac{1}{2}\varepsilon
V_3-(N^2_c-1)V_3\frac{T^4\pi^2}{45}\Omega_g-\frac{7\pi^2}{180}N_cV_3T^4
n_f\Omega_q+0(N_c^0)
\ee
where $\Omega_q$ and $\Omega_g$ are perimeter terms for  quarks and
gluons respectively,  the latter was estimated in [15] from the
adjoint Polyakov line;
 in what follows we replace
$\Omega$ by one for simplicity.

Comparing (12) and (16), $F(1)=F(2)$ at $T=T_c$, one finds in the
order $0(N_c)$, disregarding all meson and glueball contributions
\be
T_c=\left(\frac{\frac{11}{3}N_c\frac{\alpha_s<F^2>}{32\pi}}{\frac{2\pi^2}{45}
(N^2_c-1)+\frac{7\pi^2}{90}N_cn_f}\right)^{1/4}
\ee
For standard value of $G_2\equiv \frac{\alpha_s}{\pi}<F^2>=0.012
GeV^4$ [13] (note that for $n_f=0$ one should use approximately 3
times larger value of $G_2$ [13]) one has for $SU(3)$ and different
values of $n_f=0,2,4$ respectively $T_c=~240,150,134$ MeV. This
should be compared with lattice data [8] $T_c(lattice)=240,146,131$
MeV.  Agreement is quite good.  Note that at large $N_c$ one has
$T_c=0(N_c^{0})$ i.e. the resulting value of $T_c$ doesn't depend on
$N_c$ in this limit. Hadron contributions to $T_c$ are $0(N_c^{-2})$
and therefore suppressed if $T_c $ is below the Hagedorn  temperature
as it typically happens in string theory estimates  [16].

Till now we disregarded all perturbative and nonperturbative
corrections
to $F(2)$ except for magnetic condensate,
the term $\frac{1}{2}\varepsilon V_3$.
If we disregard also this term,
considering in this way only free
gas of gluons and quarks for the phase 2,
we come to the model, considered in [17].
The values of $T_c$ obtained in this
way differ from ours not much --
they are factor of $2^{1/4}=1.19$ larger, but one  immediately
encounters problems with explanation of spacial string tension,
screening masses etc.,
which are naturally are accounted for by the notion of magnetic
confinement -- nonzero values of magnetic correlators
in the  phase 2, including magnetic  condensate term,
$\frac{1}{2}\varepsilon V_3$.

 Our approximation (10) corresponding  to lowest order in $N_C$
and $g$ was too simplified when
we have put $\Omega_g=\Omega_q=1$.

Indeed NP corrections may contribute to $\Omega_g,\Omega_q$. Their
phenomenological necessity can be seen in the
measured values of $\varepsilon-3p$, [7,8], which  are seen in Fig.1
In case of $\Omega_g=\Omega_q=1$ the difference  $\varepsilon - 3p$ should
be zero, and of course higher orders in $N_C^{-1}(NP$ effects
)  and higher orders in $g$ (Perturbative effects) contribute to it.
In [15] we have tried to estimate effect of nonzero ($\Omega_g-1$),
which is $0(N_C^2)$, on the energy density and preassure. To this end
we exploit the adjoint Polyakov line and separate from  it the NP
perimeter contribution.  This can be done   if one  subtracts
properly the linear divergent perturbative contribution, specific for
Wilson and Polyakov contours.  Thus one can write [15]
\be
\Omega_g(NP)=\frac{\Omega_g(lattice)}{\Omega_g(pert)},~~
\Omega_g(pert)=exp[-\frac{N_c^2}{\beta_L}G(0)N_c]
 \ee
Substituting  this value of $\Omega_g(NP)$ into $F(2)$, eq.(16)
, one obtains $\varepsilon$ and $p$, shown in Fig.2 by the line.
 One can see a reasonable agreement between this estimate and lattice
data, which  could signify the importance of gluon perimeter contribution.
That was done for the $SU(2)$
group, since only  there exist detailed lattice data for the  adjoint
Polyakov line.

One can now estimate the  influence of nonzero $\Omega_g-1$ on $T_C$,
again for the $SU(2)$ case.

Taking  $\Omega_g(NP)$  from (18) and substituting it  into (17)
one obtains a shift of $T_C$ by a factor  of 1.07 for $n_f=2$ [18].
Similar calculations for SU(3) are now in progress [18].

\section{Spacial Wilson loops}

 In this section we derive  area law for spacial Wilson loops,
expressing spacial string tension in terms of CMC.

To this end we write $<W(C)>$ for any loop as [14]
\be
<W(C)>=exp[-\frac{g^2}{2}\int
d\sigma_{\mu\nu}(u)d\sigma_{\rho\lambda}(u')\ll
F_{\mu\nu}(u)\Phi(u,u')F_{\rho\lambda}(u')\Phi(u',u)\gg
\ee
$${\rm +\; higher\; order\; cumulants}]$$

For temporal Wilson loops, in the plane $i4, i=1,2,3,$ only color
electric fields $E_i=F_{i4}$ enter in (19), while  for spacial ones
in the plane $i,k;i,k=1,2,3$ there appear color  magnetic field
$B_i=\frac{1}{2} e_{ikl}F_{kl}$; in standard way [14] one obtains the
area law for large Wilson loops of  size $L$, $L\gg T_g^{(m)}$
($T_g^{(m)}$ is the magnetic correlation  length)
 \be
<W(C)>_{spacial}\approx exp (-\sigma_s S_{\min})
\ee
where the  spacial string tension is [10,14]
\be
\sigma_s=\frac{g^2}{2} \int d^2x\ll B_n(x)\Phi(x,0)
B_n(0)\Phi(0,x)\gg+0(<B^4>)
\ee
 and $n$ is the component normal to the plane of the contour,
 while the last term in (21) denotes contribution  of the fourth
 and higher order cumulants. On general grounds one can write for the
 integrand in (21)
 \be
 \ll B_i(x)\Phi(x,0)B_j(0)\Phi(0,x)\gg=
 \delta_{ij}(D^B(x)+D_1^B(x)+\vec{x}^2\frac{\partial D^B_1}{\partial
 x^2})-x_ix_j\frac{\partial D^B_1}{\partial x^2},
 \ee
 and only the term $D^B(x)$ enters in (21)
 \be
 \sigma_s=\frac{g^2}{2}\int d^2xD^B(x)+0(<B^4>)
 \ee
 similarly for the temporal Wilson loop in the plane $i4$ one  has
 the area law for $T<T_c$  with temporal string tension
 \be
 \sigma_E=\frac{g^2}{2} \int d^2xD^E(x)+0(<E^4>)
 \ee
 For $T=0$ due to the $0(4)$  invariance CEC and CMC coincide and
 $\sigma_E=\sigma_s$.  For $T>T_c$ in the phase (2) CEC vanish, while
 CMC change on the scale of the dilaton mass $\sim 1 GeV$, therefore
 one expects that $\sigma_s$ stays intact till the onset of the
 dimensional reduction mechanism. This expectation is confirmed by
 the lattice simulation -- $\sigma_s$ stays constant up to
 $T\approx 1.4 T_c$ [19].  Recent lattice data [19] show an increase
 of $\sigma_s$ at $T\approx 2T_c$,  for $SU(2)$ which could imply the early
 onset of dimensional reduction.

Indeed the numerical analysis  [7] shows  that at $T\geq 2T_c$
the  spacial string tension $\sigma_s$ can be well reproduced as
\be
\sqrt{\sigma_s(T)}=cg^2(T)T~~~~~~~~~~~~~~(SU_3)
\ee
which should be compared with the 3d QCD value [20]
\be
\sqrt{\sigma_s}=0.554 g^2_3~~~~~~~~~~~(SU_3)
\ee

The constant $c$ in (25) is actually expandable as a series of $g^n$,
but in the range
$2\leq T/T_c\leq 4$ it is approximately constant, $c\simeq 0.63$.
Comparison of (25) and (26) with $g^2_3=g^2(T)T$ indeed supports an early
dimensional reduction  with the coupling constant in the 2--loop
approximation
\be \frac{1}{g^2(T)}=b_0
ln(\frac{T}{\Lambda_T})^2+\frac{b_1}{b_0} lnln(\frac{T}{\Lambda_T})^2 \ee
where $b_0=\frac{11}{16\pi^2}, b_1=\frac{102}{(16\pi^2)^2}$ and the value
$\Lambda_T$ fitted to  $\sigma_s(T)$ is
\be
\Lambda_T^{\sigma}=(0.076\pm 0.013)T_c
\ee
Due to the small value of $\Lambda_T, g(T)$ is also small, $g(2T_c)\approx
\sqrt{2}$ as  compared to the $4d, T=0$ value from heavy  quarkonia, $\Lambda
\approx 200 MeV$.

Question: Why dimensional reduction sets in as early as $T=2T_c$?

\section{Screening masses of  mesons and glueballs }

This section is based on results of ref. [21].
In this section we consider the $q\bar{q}$ and $gg$ Green's
 functions $G(x,y) $ at $T>T_c$ and derive corresponding Hamiltonians
 for evolution in the spacial direction. We start with the
 Feynman--Schwinger representation [22] for $G(x,y)$, where we
 neglect for simplicity spin interaction terms \be
 G(x,y)=\int^{\infty}_0 ds \int^{\infty}_0 d\bar{s}
 e^{-K-\bar{K}}(Dz)^w_{xy}(D\bar{z})^w_{xy}<W(C)>
 \ee
 Here $K$ and $(Dz)^w_{xy}$ are defined in (10) and $<W(C)>$ in (20),
 where the contour $C$ is formed  by paths $z(\tau)$, $\bar{z}(\tau)$
 and $t\equiv x-y$ is for definiteness along the axis 3.
 Since by definition at $T>T_c$ electric correlators are zero, only
 elements $d\sigma_{\mu\nu}$ in (19) in planes 12,13 and 23
 contribute. As a result one obtains for $<W(C)>$ the form (20) with
 \be
 S_{min}=\int^t_0
 d\tau\int^1_0
 d\gamma\sqrt{\dot{w}_i^2w_k^{'2}-(\dot{w}_iw'_i)^2}
 \ee
 where only spacial components $w_i, i=1,2,3$ enter
 \be
 w_i(\tau, \gamma)= z_i(\tau)\gamma +\bar{z}_i(\tau)(1-\gamma),
 \dot{w}_i = \frac{\partial w_i}{\partial \tau}~,~~
 w'_i=\frac{\partial w_i}{\partial\gamma}
 \ee
 The form (30) is equivalent to that used before in [12] but with
 $w_4\equiv 0$.

 As a next step one can introduce " dynamical mass" $\mu,\bar{\mu}$
 similarly to [12]. We are looking for the "c.m" Hamiltonian which
 corresponds to the hyperplane where $z_3=\bar{z}_3$. Now the role of
 evolution parameter (time) is played by $z_3=\bar{z}_3=\tau$ with
 $0\leq \tau \leq t$, and we define transverse vectors
 $z_\bot(z_1,z_2),\bar{z}_\bot(\bar{z}_1,\bar{z}_2)$ and $z_4(\tau),
 \bar{z}_4(\tau)$.
 \be
  \frac{dz_3}{d\lambda}=\frac{d\tau}{d\lambda}=2\mu,~~
 \frac{d\bar{z}_3}{d\bar{\lambda}}=\frac{d\tau}{d\bar{\lambda}}=2\bar{\mu},~~
 \ee
 then $K,\bar{K}$ in (29) assume the form
 \be
 K=\frac{1}{2}\int^t_0
 d\tau[\frac{m^2_1}{\mu(\tau)}+\mu(\tau)(1+\dot{z}^2_{\bot}+\dot{z}^2_4)]
 \ee
 and the  same for $\bar{K}$ with additional bars over
 $\mu,\dot{z}_{\bot}, \dot{z}_4$.

 Performing the transformation in the functional integral (29)
 $dsDz_3(\tau)\to D\mu,~~ d\bar{s}D\bar{z}_3(\tau) \to  D\bar{\mu}$
 one has
  \be
   G(x,y)=\int
 D{\mu}D{\bar{\mu}}Dz_{\bot}
 D{\bar{z}_{\bot}}(D{z_4})^w_{xy}(D{\bar{z}_4})^w_{xy} exp (-A)
 \ee
 with the action
 \be
 A=K+\bar{K}+\sigma S_{min}
 \ee
 Note that $z_4,\bar{z}_4$ are not governed by NP dynamics and enter
 $A$ only kinematically   (through $K,\bar{K}$), and hence can be
 easily integrated out in (34) using Eq.(11) for  the 4-th
 components -- with $(x-y)_4=0$.
   One can now proceed as it was done in [
 12], i.e. one introduces auxiliarly functions $\nu(\tau,\gamma),
 \eta(\tau,\gamma)$; defines center-of-mass and relative coordinates
 $\vec{R}_{\bot}, \vec{r}_{\bot}\equiv \vec{r}$, and finally integrates out
  $\vec{R}_{\bot}$ and $\eta(\tau,\gamma)$. The  only difference from
  [12] is that now $z_4,\bar{z}_4$ do not participate in all those
  transformations. As a result one obtains \be G(x,y)\sim \int D\nu D\mu
  D\bar{\mu} Dre^{-A^{(1)}}\sum_{n,n_2}e^{-A_{n_1n_2}} \ee here \be
  A^{(1)}(\mu,\bar{\mu},\nu) = \frac{1}{2} \int^t_0
  d\tau[\frac{\vec{p}^2+m^2_1}{\mu}+
  \frac{\vec{p}^2+m^2_2}{\bar{\mu}}+\mu+\bar{\mu}
  +\sigma^2r^2\int^1_0\frac{d\gamma}{\nu}+\int^1_0\nu d\gamma+
  \ee
  $$
  +\frac{\vec{L}^2/r^2}{\mu(1-\zeta)^2+\bar{\mu}\zeta^2+\int^1_0
  d\gamma(\gamma-\zeta)^2\nu}]
  $$
  \be
  A_{n_1n_2}=\frac{1}{2}(\pi T)^2\int_0^t
  d\tau(\frac{b^2(n_1)}{\mu(\tau)}+\frac{b^2(n_2)}{\bar{\mu}(\tau)}),
  \ee
  $b(n)=2n$ for bosons and $2n+1$ for quarks. We also have introduced
  radial momentum $\vec{p}_r$, angular momentum $\vec{L}$
  \be
  \vec{p}_r^2\equiv\frac{(\vec{p}\vec{r})^2}{r^2}=
  (\frac{\mu\bar{\mu}}{\mu+\bar{\mu}})^2
  \frac{(\vec{r}\dot{\vec{r}})^2}{r^2},~~\vec{L}=\vec{r}\times
  \vec{p}
  \ee
  and
  \be
  \zeta(\tau)=\frac{\mu(\tau)+<\gamma>\int\nu
  d\gamma}{\mu+\bar{\mu}+\int\nu d\gamma}, ~~<\gamma>\equiv
  \frac{\int\gamma\nu d\gamma}{\int\nu d\gamma}
  \ee

Let us  define
the Hamiltonian $H$ for the given action $A=A^{(1)}+A_{n_1n_2}$ in (36),
integrating over $D{\nu},D{\mu},D{\bar{\mu}}$ around the extremum of
$A$ ( this is an exact procedure in the limit $t\to \infty$).
For the extremal values of auxiliary fields one has
$$
\frac{\vec{p}^2+m^2_1+(b(n_1)\pi T)^2}{\mu^2(\tau)} = 1
-\frac{l(l+1)}{\vec{r}^2}(\frac{(1-\zeta)^2}{a^2_3}-\frac{1}{\mu^2})$$
 \be
\frac{\vec{p}^2+m^2_2+(b(n_2)\pi T)^2}{\bar{\mu}^2(\tau)} = 1
-\frac{l(l+1)}{\vec{r}^2}(\frac{\zeta^2}{a^2_3}-\frac{1}{\bar{\mu}^2})
\ee
$$
\frac{\sigma^2}{\nu^2(\tau,\gamma)}\vec{r}^2=
1-\frac{l(l+1)}{\vec{r}^2}\frac{(\gamma-\zeta)^2}{a^2_3}$$
where
$a_3=\mu(1-\zeta)^2+\bar{\mu}\zeta^2+\int
d\gamma(\gamma-\zeta)^2\nu$ and $\zeta$ is
defined  by eq.(40).

After the substitution of these extremal
values into the path integral Hamiltonian
   \be
    G(x,y) = <x|\sum_{n_1n_2}e^{-H_{n_1n_2}t}|y>
    \ee
one has to construct (performing proper Weil
ordering  ) the operator Hamiltonian
acting on the wave functions.

Consider for simplicity the case $\vec{L}=0$, then from (37-38) one
obtains

 \be
  H_{n_1n_2}= \sqrt{\vec{p}^2+m^2_1+(b(n_1)\pi T)^2}+
\sqrt{\vec{p}^2+m^2_2+(b(n_2)\pi T)^2}+ \sigma r
\ee
Here $\vec{p}=
\frac{1}{i}\frac{\partial}{\partial\vec{r}}$ and
$\vec{r}=\vec{r}_{\bot}$ is a $2d$ vector,
$\vec{r}=(r_1,r_2);~m_1,m_2$ -- current masses of
quark and antiquark, for $gg$ system $m_1=m_2=0$
and $\sigma=\sigma_{adj}=\frac{9}{4}\sigma$ .

Eigenvalues and eigenfunctions of $H_{n_1n_2}$
\be
H_{n_1n_2}\psi(r)=M(n_1,n_2)\psi(r)
\ee
define the so--called screening masses and corresponding
wave--functions, which have been measured in lattice calculations
[5,6,23].

The lowest mass sector for mesons is given by $H_{00}$, where for
$n_1=n_2=0$ one has $b(n_1)=b(n_2)=1$ in (43). For light quarks one
can put $m_1=m_2=0$ and expand at large $T$ square roots in (43) to
obtain
\be
 H_{00}\approx 2\pi T+ \frac{\vec{p}^2}{\pi T}+ \sigma r
,~~ M_{00}\approx 2\pi T + \varepsilon (T)
\ee
 where $\varepsilon(T)=\frac{(\sigma_s(T))^{2/3}}{(\pi T)^{1/3}}a $
 and $a\simeq 1.74$ is the eigenvalue of the 2-D dimensionless
 Schr\"{o}dinger equation.

 Assuming the parametrization $\sqrt{\sigma_s} =cg^2(T)T$ with
 $c=0.369$ and scaling behaviour of $g^2(T)$ [19] one has
 $M_{00}\approx 2\pi T+0((\ln(T/\Lambda_T))^{-4/3}T)$ tending to
 twice the lowest Matsubara frequency. (This limit corresponds to the free
quarks, propagating
 perturbativly in the space--time with the imposed antiperiodic
 boundary conditions along the 4th axis).

Eq. (45) coincides with that proposed in [24], where also numerical
study was done of $M_{00}$ and $\psi_{00}(r)$ .
Our calculations of Eq.(44-45) agree with [24] and are
presented in Fig.1 together with lattice
calculations of $\psi_{00}(r)$ for $\rho$ -- meson
[23].
 The values of $M_{00}(T)$ found on
the lattice [3,4] are compared with our results in Fig.4.  Note, that
our $M_{00}$ (45) does not contain perimeter corrections which are
significant.  Therefore one has to add the meson
 constant to $M_{00}$ to compare with lattice data.  We
disregard spin--dependent and one-gluon-exchange (OGE) interaction
here for lack of space.
It is known [24] however, that OGE is not much important at around
$T\approx 2T_C$.

We note, that the lowest meson screening mass  appear also in Yukawa
type exchanges  between quark lines and can be  compared with the
corresponding Yukawa potentials.

 For the $gg$ system the lowest
mass sector is given by $b(n=0)=0$, and one has from (43)
\be
H_{00}(gg)=2|\vec{p}|+\sigma_{adj}r \ee
 Note that $T$ does not enter
the kinetic terms of (46).

To calculate with (46) one can use  the approximation in (37) of
$\tau$--independent $\mu$ [12], which leads to the operator
$(\mu=\bar{\mu})$
 \be h(\mu)=\mu+\frac{\vec{p}^2}{\mu}+\sigma_{adj}r
\ee
The eigenvalue $E(\mu)$ of $h(\mu)$ should be minimized with respect to $\mu$
and the result $E(\mu_0)$ is known to yield eigenvalue of (46)
within few percent accuracy [25]. The values $E(\mu_0)\approx
M_{gg}$ thus found are  presented  in Fig.4. The
corresponding wave functions $\psi_{00}(r)$ are given in Fig.3.
These data can be compared with the glueball screening masses, found on the
lattice in [6].

Another point of comparison is Polyakov line correlator
\be
P(R)=<\Omega(R)\Omega(0)>-<\Omega(R)><\Omega(0)>= exp
(-V(R)/T)-1
\ee
 with $$V(R)=\frac{exp(-\mu
R)}{R^{\alpha}}~,~~\Omega(R)=\frac{1}{N_C} tr P exp ig\int^{\beta}_0
A_4dz_4 $$
 It is easy to understand that $\mu=M_{gg}$ and one can
compare our results for
glueball screening mass (GSM) $M_{gg}(T)$ with the corresponding
lattice data for  $\mu$ in Fig.4.

 Taking  again into account an unknown constant perimeter correction to our
 values of $M_{gg}$, one can see a reasonable qualitative behaviour.

In addition to the purely nonperturbative source of GSM described above
there is a  competing mechanism -- the perturbative formation of the electric
Debye screening
mass $m_{el}(T)=gT(\frac{N_c}{3}+\frac{n_f}{6})$ for each gluon
 with $g(T)$  given by the temperature dependence of
$\sqrt{\sigma_s(T)}=c g^2(T)T$ [7,19]. Therefore for large $T$, where
$m_{el}$ is essential, one should rather use instead of (46) another
Hamiltonian, which is obtained from (43) replacing $m_1=m_2=m_{el}(T)$ \be
\tilde{H}_{00}(gg)=2\sqrt{\vec{p}^2+m^2_{el}(T)}+\sigma_{adj}(T)r
\ee
It turnes out that up to the temperatures  $T\leq 2 T_c$ one can
consider the effect of   $m_{el}(T)$  as a small
correction to the eigenvalue of (46) $E$
\be
M_{gg}=  E+\delta E\approx 4(\frac{a}{3})^{3/4}\cdot \frac{3}{2} c g^2
(T)T+\frac{T}{\frac{3}{2}c(\frac{a}{3})^{3/4} },
\ee
and that gives us the right to treat GSM in the temperature region
concerned nonperturbatively.

We note that the  transition between these two regimes (with  the
dominance of nonperturbative and then perturbative dynamics) is
smooth,  with GSM tending to twice the Debye mass at large $T$.

\section{Infrared catastrophe in Hot QCD}

The infrared (IR) catastrophe in QCD was indentified within the  hot
perturbative series in [3].  For our purposes we shall explain it in the
language of Feynman diagrams in the  configuration space. Consider the
Feynman diagram for the thermodynamic potential in the order $g^n$, shown in
Fig.5. It can be written as
\be
J^{(n)}=\prod^n_{i=1} d^4x^{(i)} \Gamma^{(i)}\prod_{i,j} G(x^{(i)},x^{(j)})
\ee
where $G_{ij}$ is the  $x$-- space gluon Green's function connecting
vertices $i$ and $j$, and $\Gamma^{(i)}$ is the three--gluon vertex
containing one derivative, $\partial/\partial x^{(i)}$. We suppress
all color and Lorentz indices for the sake of brevity and will only
count the powers of $x^{(k)}$ to judge the IR convergence of
$J^{(n)}$.

At $T>0$ using Matsubara formalism one has to replace
$$
d^4x\to T^{-1}d^3 x
$$
\be
G(x,y)\to (
\partial^2)^{-1}_{xy}=\sum_{k=0,\pm1,...}\int \frac{T d^3p}{(2\pi)^3}
\frac{e^{-i\vec{p}(\vec{x}-\vec{y})-i2\pi kT(x_4-y_4)}}{(\vec{p}^2+
(2\pi kT)^2)}
\ee
For large $T$ the only mode propagating  over distances larger that $1/T$,
as follows from (52)  is the mode with $k=0$. Exactly this mode causes IR
catastrophe; we have for it
\be
G(x,y)\sim \frac{T}{4\pi |\vec{x}-\vec{y}|},~~ |\vec{x}-\vec{y}|\gg 1/T
\ee
Now for the diagram of the type of Fig.(5) one has  for a given number of
vertices $n$ total number of propagators $\frac{3n}{2}$, so that one can
write (remembering that one vertex yields overall volume)
\be
J^{(n)}\sim g^nV_3T^{-n}\frac{(d^3x)^{n-1}\cdot
T^{\frac{3n}{2}}}{x^{\frac{5n}{2}}}\sim
\ee
$$
\sim g^nV_3T^{\frac{n}{2}}\int (x^{\frac{n}{2}-3})
$$
where the  the symbol $\int(x^k)$ denotes the overall  power of $x^{(i)}$ in
the integrand including those of $d^3x^{(i)}$.
One can immediately see that for $n\geq 6$ the diagram $J^{(n)}$ diverges at
large $x$.

We shall now  give simple arguments which show that accounting for the
nonperturbative background,  the perturbation series for $F$ behaves well
and IR catastrophe disappears.

To this end we consider  the perturbation theory given by Eqs. (1-11).

The diagram of Fig.(5) goes over into the equivalent diagram of the order
$g^n$, where vertices and propagators are now taken in the NP background
field.
\be
G(x,y)\to \left\{
\begin{array}{ll}
(-D^2)^{-1}_{xy},& {\rm ~Eq.(10)~for~the~ghost}\\
(W)^{-1}_{xy}=(-D^2-2gF)^{-1}_{xy},& {\rm~for~ the ~gluon}
\end{array}
\right.
\ee
 $$
 \Gamma^{(i)}\sim\partial/\partial x^{(i)}_{\mu}\to
 \frac{\partial}{\partial x_{\mu}^{(i)}}-ig B_{\mu}(x^{(i)})
 $$
 The expression for the diagram, $\tilde{J}^{(n)}$ looks the same as (53)
 with eq.(10), for propagators, but now two major changes appear:

 i) All phase factors  $\phi (x,y)$ of gluon (ghost) Green's function  and
 vertixes $\Gamma$ assemble into gauge--invariant combinations of
 Wilson loops and covariant derivatives, so that the diagram is
 gauge--invariant.

 ii) Averaging over background fields as in (2) brings about Wilson--loop
 averages, which are subject to area law for large loops (as can be seen
 from cluster expansion [14]  or lattice calculations [19]).

 Therefore it is highly plausible that large distance behaviour of the total
 diagram is cut -- off at large distances $|x^{(i)}-x^{(j)}|$ by the Wilson
 loops; by dimentional arguments it reduces to saying that the divergent
 integral
 $\int (x^{\frac{n}{2}-3}),~~(n\geq 6)$ should be replaced  by
 $\mu^{3-n/2}$, with $\mu\approx \sqrt{\sigma_s}$.

 Correspondingly Eq.(54)  is replaced in the framework of the NP background
 perturbation theory by
 \be
 \tilde{J}^{(n)}\sim g^nV_3T^{n/2}(\sqrt{\sigma_s})^{(3-n/2)}
 \ee
 Now at high $T, T\gg 2T_c$, as we discussed in chapter 4, the dimensional
 reduction mechanism is at work, which yields [7,19]
 $\sqrt{\sigma_s(T)}=cg^2(T)T$. Inserting this into (56) one gets
 \be
 \tilde{J}^{(n)}\sim
 V_3T^3(cg^2(T))^3(\frac{g}{c})^{n/2}
 \ee

 Thus all diagrams with $n\geq 6$ are not IR diverging due to the
 Spatial string tension $\sigma_s$ and their contribution to the free
 energy density can be written in the same form as in [26] \be
 \frac{f(T)}{T}=\frac{\Sigma\tilde{J}^{(n)}}{V_3}=(g^2T)^3f_G+...
 \ee
 where $f_G=\sum_{n\geq 6} a_n (\frac{g}{c})\frac{n-6}{2}$ is
 given by the sum of diagrams with $n\geq 6$, and can be easily
 computed since $g(T)$, Eq.(27) is small for large $T$. Note however,
 that $c=c(g)$ as discussed, below Eq.(26).

 Since our conclusion about the absence of the IR catastrophe is crucial for
 the theory, we present  now another argument on convergence of
 $\tilde{J}^{(n)}$.  To this end let us consider behaviour of the diagram,
 Fig.5., when one of the vertices, say $i=1,$ is far from all others. The
 diagram is actually longdistance limit of $3g$ glueball Green's function,
 since 3 gluons are produced at the vertex 1 in the gauge--invariant state
 with total angular momentum zero (this vertex is given by $L_6(a^3)$ term
 in (4)).

 It is clear, that at large $|x^{(1)}-X|$, where $X$ stands for the assembly
 of all other
 vertices, the diagram behaves as
 \be
 \tilde{J}^{(n)}(x^{(i)},x^{(2)}...)\sim e^{-M_{3g}(0^+)|x^{(1)}-X|}
 \ee
 where $M_{3g}(0^+)$is the mass of the 3--gluon glueball in 3d, one expects
 that $M_{3g}(0^+)>1GeV$.

 Now $M_{3g}(0^+)\sim \sqrt{\sigma_s} const$, where $const$ is
 some number of order of one. Thus again we obtain a cut--off in the
 $x$--space, which makes all integrals convergent and the estimate
 (57) remains true.

 We note that nonperturbative background has created a situation equivalent
 (in the sense of convergence) to the appearance of the  magnetic mass,
 \be
 M_{magn}\approx \sqrt{\sigma_s}\sim g^2T
 \ee
 Note however, that $M_{magn}$ appears in a gauge--invariant manner, and not
 in $\prod_{ij}$ -- self energy part of gluon. Rather it appears as in the
 constituent gluon model; if one defines $M_{gg}=2M_{magn}$ (cf.Eq.(50)),
 then $M_{3g}\approx 3 M_{magn}$, and each gluon propagator contributes in
 $x$-space assymptotics $G(x,y)\sim e^{-M_{magn}|x-y|}$.
 This asymptotics reproduces immediately (59) and (56). Also the Polyakov
 line correlator (48) containing the exchange of two gluon lines, decreases
 at large distances with screening mass $\mu=M_{gg}\approx 2M_{magn}$.

\section{Conclusions}

We have presented the formalism of QCD at $T>0$, which contains NP field in
the background, in the form of gauge--invariant correlators and perturbative
expansion in this background.

It is shown how the Feynman--Schwinger representation  (FSR) which
proved to be very useful for $T=0$, is modified for $T>0$. In
particular, all gluon and quark propagators are written as a sum of
path integrals with winding paths in the 4th direction.

Using this formalism the structure of vacuum at $T<T_c$ and $T>T_c$ as well
as the nature of deconfinement transition is clarified and $T_c$ is computed
in good agreement with lattice data.  Moreover the spacial area law is shown
to follow  naturally from the structure of the vacuum at $T\geq T_c$. Using
FSR the Hamiltonian for screening states of mesons and glueballs is derived,
and screening masses and wave functions for lowest meson and glueball states
are numerically computed.

Finally, it is demonstrated how the Infrared Catastrophe is eliminated due
to the NP background and estimates  of higher order diagrams for the free
energy are given. All this exemplifies NP background field theory and in
particular perturbation theory as a powerful instrument both for
$T<T_c$ and for $T>T_c$.

The author acknowledges the financial support from the Russian Fund
for Fundamental Research, Grant 93-02-14937.

  \newpage

  \begin{center}
  {\bf Figure captions}\\

  \end{center}

  Fig. 1. The intraction measure, ($\varepsilon-3p/T^4$  as a
  function of temperature, from ref. [7].\\

  Fig. 2.  The normalized pressure $p/p_{SB}$ and energy density
  $\varepsilon/\varepsilon_{SB}$ calculated from (16), (18) (p.back.,
  e.back) and lattice calculations (p.latt., e.latt) as
  functions of $ T/\Lambda_L$, from ref. [15].\\

  Fig. 3. The $\rho$-- meson (solid line) and glueball (dashed line)
  wave. functions for lowest Matsubara frequencies $vs~r/a (a=0.23
  fm)$ for $T=210 MeV$ from ref. [21]. The lattice data are from ref
  [23] for the same $T$.\\

  Fig. 4. The screening masses for mesons (solid line) and glueballs
  (dashed line) as functions of temperature from ref. [21]. The
  lattice data are from ref. [5] (squares) and  ref. [6]
  (triangle).\\

  Fig. 5. The typical IR divergent diagram of hot QCD in the order
  $g^n$.

  \newpage

  \Large

  Fig.1,~~~
  Fig.2,~~~
  Fig.3,~~~
  Fig.4,~~~
  Fig.5,\\
       \\
       \\

  $x_1$,~~
  $x_2$,~~
  $x_3$,~~
  $x_4$,~~
  $x_{n-1}$,~~
  $x_n$
   \end{document}